\documentclass{emulateapj}

\usepackage{apjfonts}

\newcommand{\rmn}[1]{{\mathrm{#1}}} 
\newcommand{\etal}{{et\,al.}}  
\newcommand{\fig}[1]{Figure~\ref{fig:#1}}
\newcommand{\tab}[1]{Table~\ref{tab:#1}}
\newcommand{\subsect}[1]{Section~\ref{subsect:#1}}
\newcommand{\HI}{\hbox{H~$\scriptstyle\rm I\ $}}

\shorttitle{$U$-band dropouts in the HUDF parallels}
\shortauthors{Wadadekar et. al}
\submitted{accepted by the Astronomical Journal}
\begin{document}

\title{Faint $U$-band dropouts in the WFPC2 parallels of the Hubble
Ultra Deep Field} \author{Yogesh Wadadekar\altaffilmark{1}, Stefano
Casertano\altaffilmark{1} and Duilia de Mello\altaffilmark{2,3,4}}
\altaffiltext{1}{Space Telescope Science Institute, 3700 San Martin
Drive, Baltimore, MD 21218} \altaffiltext{2}{Observational Cosmology
Laboratory, Code 665, Goddard Space Flight Center, Greenbelt, MD
20771} \altaffiltext{3}{Catholic University of America Washington, DC
20064}

\altaffiltext{4}{Johns Hopkins University, Baltimore, MD 21218}

\begin{abstract}

We combine data from the extremely deep Hubble Space Telescope $U$
 (F300W) image obtained using WFPC2 as part of the parallel
 observations of the Hubble Ultra Deep Field campaign, with $BVi$
 images from the Great Observatories Origins Deep Survey (GOODS) to
 identify a sample of Lyman break galaxies in the redshift range $2.0
 \lesssim z \lesssim 3.5$.  We use recent stellar population synthesis
 models with a wide variety of ages, metallicities, redshifts, and
 dust content, and a detailed representation of the \HI cosmic opacity
 as a function of redshift to model the colors of galaxies in our
 combination of WFPC2/ACS filters.  Using these models, we derive
 improved color selection criteria that provide a clean selection of
 relatively unobscured, star forming galaxies in this redshift
 range. Our WFPC2/F300W image is the deepest image ever obtained at
 that wavelength.  The $10\sigma$ limiting magnitude measured over 0.2
 arcsec$^2$ is 27.5 magnitudes in the WFPC2/F300W image, about 0.5
 magnitudes deeper than the F300W image in the Hubble Deep Field
 (HDF)-N. This extra depth relative to the HDFs allows us to directly
 probe the luminosity function about 0.5 magnitudes deeper than the
 depth accessible with the HDF data along an independent line of
 sight. Our sample of star-forming galaxies with $2.0 \lesssim z
 \lesssim 3.5$ includes 125 objects, the majority of which show clumpy
 morphologies.  We measure a star formation rate density of 0.18
 $M_\odot \rm yr^{-1} Mpc^{-3}$, marginally higher than the value
 measured for the Hubble Deep Fields.

\end{abstract}
  

\section{Introduction}

Great effort has been expended in the last decade to identify galaxies
at high redshift.  It is hoped that measurements of their luminosity
and color evolution will place constraints on the star formation
history of the universe and indirectly on the formation of large scale
structure.  In this search for high redshift galaxies, the Lyman
continuum break technique has been successfully applied to detect
galaxies at cosmological distances.

A characteristic feature in the UV spectrum of star forming galaxies
is the Lyman break, the sharp continuum discontinuity at 912 \AA.  The
break is caused by absorption of UV photons with wavelength $\lambda <
912$ \AA \ in the outer atmospheres of massive stars that produce such
photons, in the interstellar \HI gas, and by intervening \HI gas along
the line of sight from the star forming galaxy to the observer.  At a
redshift of $z \sim 3$, the Lyman break is redshifted longward of the
$U$ filter, so that galaxies have very low (or undetectable) flux in $
U $.  If the galaxy is starforming, it will have a high flux redward
of the Lyman break, causing it to be bright in observed $B$ and blue
in colors such as $B-V$ or $V-I$.  A color selection that identifies
``$U$ dropouts'', i.e., galaxies that are blue in $B-V$ and very faint
in $U$, would then preferentially select actively starforming galaxies
in the redshift range $ 2 \lesssim z \lesssim 3.5 $. The same
principle can be used to identify galaxies at higher redshifts by
moving to a redder set of filters.  For example, the $B$ dropouts
which have $ z \sim 4.5$ can be effectively identified using $BVI$
data.

Several research groups have identified and studied $U$ dropout
galaxies (see Giavalisco (2002) for a review).  Ground based studies,
mostly carried out with the Keck telescope (e.g., Steidel {\etal} 2003
and references therein), have studied samples with deep imaging and
spectroscopy over relatively large areas of sky (typically about 1000
square arcmin), to a typical depth of $R_{AB}=25.5$ largely set by the
spectroscopic follow-up.  These studies were complemented by samples
identified using the Hubble Deep Fields, two deep imaging surveys with
the WFPC2 instrument aboard the Hubble Space Telescope (HST) (Williams
{\etal} 1996; Casertano {\etal} 2000, hereafter C00).  The Hubble Deep
Fields were used to identify and study $U$ dropout galaxies (Madau
{\etal} 1996; Dickinson (1998); C00) to much deeper limiting
magnitudes ($R_{AB} \sim 28.0$), but over a solid angle about two
orders of magnitude smaller than the ground-based observations. The
HDF samples were too faint for complete spectroscopic follow-up.  The
precision of HST photometry also ensures small random errors on color
measurements.  Another important advantage of the HST is its
unprecedented angular resolution that allows for a study of the main
morphological characteristics of the $U$ dropout galaxies, even at
high redshift.  However, the small solid angle that they probe makes
them susceptible to statistical fluctuations caused by small number
statistics, galaxy clustering and field-to-field variations.  The
amplitude of these fluctuations can be estimated by observing multiple
fields separated by a large angular distance.  Presently, the two
Hubble Deep Fields provide the only two such fields on the sky.

The scientific success of the WFPC2 Hubble Deep fields motivated an
even deeper set of observations with the Wide Field Channel of the
Advanced Camera for Surveys (ACS) installed on the HST in March 2002.
The Hubble Ultra Deep Field (HUDF; Beckwith {\etal} 2006) covered an
area about twice that of each of the original deep fields and reaches
about 1.5 magnitudes deeper in the $i$ band.

The ACS Wide Field Channel has been optimized for visible and near-IR
throughput and has very poor UV sensitivity.  For this reason, the
HUDF ACS observations were restricted to the $BViz$ (F435W, F606W,
F775W and F850LP) filters; $U$ band data were not obtained.  The HUDF
thus provides the deepest available sample of $B$, $V$, and $i$
dropout galaxies, but a study of $U$ dropouts is not possible.
However, in parallel with the HUDF ACS and NICMOS prime observations,
data were gathered with all the other HST instruments in parallel
mode.  As part of these parallel observations, the WFPC2 instrument
observed two fields using the F300W filter.  Both these fields are
considerably deeper than the HDF F300W observations.  One of them
overlaps with the coverage of the ACS GOODS-South field (Giavalisco
{\etal} 2004) and thus has moderately deep $BViz$ data from that
survey.  In this paper, we combine data from the GOODS survey with the
WFPC2 parallel observations of the HUDF to identify a new sample of
$U$ band dropout galaxies.  Our deeper $U$-band observations, relative
to the Hubble Deep Fields, allow us to directly explore a region of
the luminosity function of $U$ dropout galaxies that has never been
directly probed before.

Throughout this paper, unless otherwise stated, we use the standard
concordance cosmology with $\Omega_M=0.3$, $\Omega_{\Lambda}=0.7$~and
$h_{100}=0.7$.  For magnitudes, we use the AB system of Oke and Gunn
(1983). For convenience, we use the designation $U$, $B$, $V$ and $i$
while referring to the WFPC2/F300W, ACS/F435W, ACS/F606W and ACS/F775W
filters respectively.

This paper is organized as follows: in Section 2 we describe our data
and data processing procedures. In Section 3 we describe the
simulations we carried out to identify the appropriate color selection
criteria for our dropout candidates and how we applied the criteria to
identify our dropout sample. Section 4 describes the dropout
morphology and Section 5 contains a discussion of dropout statistics.

\section{The data}

\begin{figure*}
\plotone{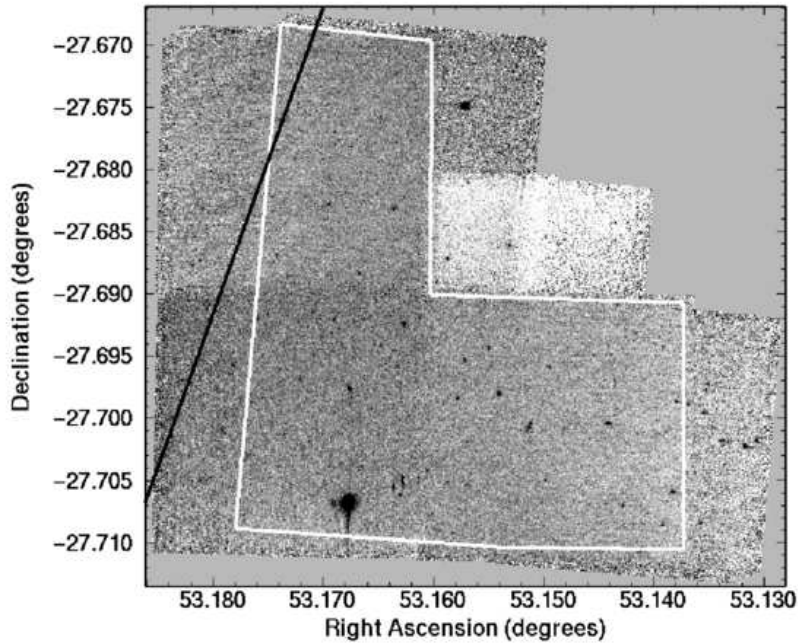} 
\caption{The final drizzled image in F300W obtained by combining 409
individual exposures. The white polygon encloses the area over which
the wide field chips of all datasets overlap to give a total exposure time of 323.1 ksec,
outside this area effective exposure is between 25\% and 75\% of the
full depth. Portions of the image above and to the left of the black line lie outside the
GOODS area and are not useful for identifying $U$ dropout
galaxies.}\label{fig:final}
\end{figure*}

In this paper we analyze the HUDF/WFPC2 (F300W) parallels which
overlaps substantially over the GOODS-S area i.e. the Epoch1-Orient
310/314 image (see Fig. 1 of de Mello et al. (2006) for a finding
chart of this area relative to the GOODS-S field).  The other WFPC2
parallel field (Epoch 2-Orient 40/44) overlaps with the Galaxy
Evolution From Morphology and SEDs (GEMS; Rix {\etal} 2004) survey
area; because of the lack of $B$-band data, this region cannot be
studied in the same way, and is not considered further in this paper.
Each field includes several hundred exposures with a total exposure
time of 323.1 ks and 278.9 ks respectively.  Because of small changes
in the position angle of HST during each observing campaign, data at
each epoch cover 6.52 square arcmin, with 4.02 square arcmin at the
full nominal depth (the three wide field chips from all datasets
overlapping), and the remaining 2.50 between 25\% and 75\% of the
observing time.  About 87\% (5.67 square arcmin) of the F300W image
overlaps with the GOODS-S footprint, including almost the entire full
depth area (see \fig{final}).

\subsection{Data Processing}

The data processing was carried out with the techniques developed for
the WFPC2 Archival Parallels Project (APPP; Wadadekar {\etal} 2006),
based on the drizzle approach (Fruchter \& Hook 2002).  We constructed
a drizzled image with a pixel scale of 0.06 arcsec/pixel. The various
procedures required to obtain a final combined image have been
described in de Mello et al. (2006). For completeness, we repeat the
data processing steps here.

For Epoch 1, a total of 409 WFPC2/F300W parallel images, with exposure
times ranging from 700 seconds to 900 seconds were obtained in
parallel with the prime ACS observations. Each of the datasets was
obtained at one of two orientations of the telescope: (i) 304 images
were obtained at Orient 314 and (ii) 105 images were obtained at
Orient 310.

We downloaded all 409 datasets from the HST data archive along with
the corresponding data quality files and flat field images. These
datasets were all downloaded in a single request to the archive to
ensure that the ``on the fly reprocessing'' (OTFR) procedure was
identical for all the images.  World Coordinate System (WCS)
coordinates listed in the FITS header of WFPC2 images may be
inaccurate by as much as an arcsecond, depending on the details of the
HST guide star acquisition process. This inaccuracy can lead to
relative offsets between images that contribute to the drizzle stack,
as well as errors in the absolute positioning of the image on the sky.

The processing pipeline of the APPP registers images relative to each
other using an automated procedure that employs matched source lists
to compute the offsets. We modified this procedure by measuring the
centroid positions of 4 predetermined stars in every image with
respect to an arbitrarily chosen reference image. These stars are the
only sources that are detectable in each individual exposure in our
drizzle stack. The nominal WCS of each individual image in the drizzle
stack was updated to reflect the improved relative alignment with
respect to the reference image; after this WCS update the images were
well registered with respect to the reference image. Given that we
only had 4 stars to fit to, we were unable to determine independent
rotation corrections to a sufficiently high precision, and therefore
left unchanged the orientation of each individual image, correcting
only for shifts in the position of the stars.

Our procedure also accounted for the shifts in the relative positions
of the WFPC2 detector as a function of time using the model of
Casertano \& Wiggs (2001). Geometric distortion coefficients
determined by Kozhurina-Platais {\etal} (2003) for the F300W filter
were incorporated into the procedure. Cosmic ray rejection was achieved
using the procedure outlined by Fruchter \& Hook (2002).

The processing was carried out in two stages. In the first stage,
images with near 100\% overlap were grouped together in groups of 20
images each. Each group was processed separately. Such separate
processing in groups were necessitated by memory limitations on our
computer.  The drizzled image stack of 20 images was used to identify
cosmic rays and store that information in a cosmic ray flag image for
each of the 20 images combined. This step was repeated for all the
groups. In the second stage, all 409 images were drizzled through to a
final image using the cosmic ray flag images to reject cosmic rays.

Offsets between our final drizzled image and the GOODS images were
measured by matching sources in our image with the corresponding
sources in the GOODS data, which were binned from their original scale
of 0.03 arcsec/pixel to 0.06 arcsec/pixel. Once the offsets between
the drizzled WFPC2 image and the GOODS images had been measured, all
409 images were drizzled through again taking the offsets into
account, so that the final image was accurately aligned with the
binned GOODS images. This final image had an identical World
Coordinate System (WCS) as the binned GOODS images; this means that
all sources have identical pixel coordinates in all filters and can
thus be used for accurate dual mode photometry, where the image in one
filter is first used for detection and subsequently photometry is
carried out in all filters using the segmentation map of the detection
image.  The residual scatter in source registration between the GOODS
images and the WFPC2 image is random (20 milliarcsec r.m.s.)  and is
likely caused by the mismatch in the centroids of sources between the
$U$ and redder bands, which is probably due to the fact that the
regions of peak UV emission in galaxies probed by the $U$ band image
may not coincide with the peaks of light emission at redder
wavelengths, leading to random offsets in centroid positions.

We  weighted each drizzled dataset by its inverse variance map during
the drizzle process. This map was computed according to the
prescription of C00. This computation of the variance takes into
account contributions to the noise from the sky background (as
modulated by the flat field), dark current and read
noise. Contributions to shot noise from sources are not included. Each
weight map is combined with masks that exclude (i.e., set to zero
weight) pixels that are flagged as bad in the data quality files, and
is then used to weight the combination of images in the drizzle
process.  The drizzle process also produces a weight map as one of its
outputs. The weight map is an accurate measure of the effective
exposure (and therefore depth) that is reached at any particular
location in the image after correctly accounting for missing data due
to cosmic rays, hot pixels etc. in each image that is being combined.

The WFPC2 CCDs have a small but significant charge transfer efficiency
problem (CTE) which causes some signal to be lost when charge is
transferred down the chip during readout (Heyer, Biretta, et
al. 2004). The extent of the CTE problem is a function of target
counts, background light and epoch. Low background images (such as
those in the F300W filter) at recent epochs are more severely
affected. Not only sources, but also cosmic rays leave a significant
CTE trail. We attempted to flag the CTE trails left by cosmic rays in
the following manner: if a pixel was flagged as a cosmic ray, adjacent
pixels in the direction of readout (along the Y-axis of the chip) were
also flagged as cosmic-ray affected. The number of pixels flagged
depends on the position of the cosmic ray on the CCD, with more pixels
flagged for higher row numbers. Only one pixel was flagged for rows 1
through 100, 2 pixels were flagged for rows 101 through 200 and so
on. With this approach, we were able to eliminate most of the
artifacts caused by cosmic rays in the final drizzled image.

The 10$\sigma$ limiting magnitude measured over 0.2 arcsec$^{2}$ is
27.5 magnitudes over the $\sim$4 square arcmin field that has all 409
datasets overlapping, which is about 0.5 magnitudes deeper than the
F300W image in the HDF-N and 0.7 magnitudes deeper than that in the
HDF-S. Somewhat lower depth is reached in the regions where not all
datasets overlap.

\section{Modeling the colors of high redshift galaxies}

\subsection{Evolutionary population synthesis models}

In order to identify galaxies in the redshift range of interest,
candidates first need to be identified in an appropriate color-color
space of the available data. Such a ``selection function'' may be
affected by many factors: the particular color criteria adopted,
the intrinsic dispersion in the UV spectral properties of the star
forming galaxies (as influenced by differences in metallicity, stellar
initial mass function, star formation history and dust properties),
the cosmic variance along different lines of sight and  the
photometric measurement errors. This selection function can be
estimated from models or can be measured directly, by determining
enough spectroscopic redshifts to measure it empirically. Empirical
determination for space based dropout samples is precluded for at
least two reasons. 1. faintness of the candidates makes spectroscopy
difficult even with 10 m class telescopes and 2. clustering of Lyman
break galaxies necessitates measurements over many independent
sightlines to average over the effects of large scale structure. The
two deep fields provide the only currently available sightlines for
HST data. We therefore need to use a model to obtain the color
selection criteria that isolate high redshift star forming galaxies.

The colors of galaxies as a function of age, metallicity, stellar
initial mass function and star formation history can be effectively
modeled using the evolutionary population synthesis technique (see
Bruzual \& Charlot 2003, hereafter BC03, and references
therein). Assumptions about the time evolution of these parameters
allow one to compute the age dependent distribution of stars in the
Hertzsprung-Russell diagram, from which the integrated spectrum for
the entire galaxy can be obtained. The models exploit the property
that stellar populations with an arbitrary star formation history can
be thought of as the sum of a series of instantaneous starbursts
(delta functions), referred to as simple stellar populations
(SSPs). The spectral energy distribution at time $t$ of a stellar
population characterized by a star formation rate $\psi(t)$ and a
metal-enrichment law $\zeta(t)$ can be written (following, e.g.,
Tinsley 1980)

\begin{equation}
F_\lambda(t) =
\int_0^t\,\psi(t-t')\,S_\lambda\left[t',\zeta(t-t')\right]\, dt'\,,
\label{convol}
\end{equation}
where $S_\lambda\left[t',\zeta(t-t')\right]$ is the power radiated per
unit wavelength per unit initial mass by an SSP of age $t'$ and
metallicity $\zeta(t-t')$. The above expression assumes that the
initial mass function (IMF) does not change with time.

The largest source of uncertainty in the spectral synthesis technique
arises from the relatively poorly understood advanced phases of
stellar evolution, such as the supergiant and asymptotic giant branch
(AGB) phases (BC03).

In this paper, we use predictions from the BC03 model for the purpose
of defining the color selection criteria for $U$ dropout galaxies for
the combination of ACS and WFPC2 filters specific to our data. We
chose to use the BC03 model, because it is a recent model that
provides high resolution spectra for our wavelength range of
interest. The code implementing the model prescriptions, named
GALAXYEV, is publicly available from the authors, thus providing a
means of making our simulations reproducible by other researchers.

We summarize here the main features of the BC03 model that are
relevant for our purposes. The model predicts the spectral evolution
of stellar populations of different metallicities and ages between $1
\times 10^5$ and $2 \times 10^{10}$ years at a resolution of 3 \AA\
FWHM over the whole wavelength range from 3200-9500 \AA. These
predictions are based on a new library of observed stellar spectra
compiled by Le Borgne (2003) called STELIB.  Predictions over a wider
wavelength range but at lower resolution are also available. The
Padova 1994 stellar evolution prescription and the BaSel 3.1 spectral
calibrations are used.  Two choices of IMF are available: the standard
Salpeter (1955) IMF and the Chabrier (2003) IMF with lower and upper
mass cutoffs of 0.1 $M_{\odot}$ and 100 $M_{\odot}$ respectively. Each
model SSP is normalized to a total mass of 1 $M_{\odot}$ in stars. The
spectra are computed at 221 unequally spaced SSP ages from 0 to 20
Gyr. Metallicity ranges from 0.005 to 2.5 $Z_{\odot}$.

The authors of BC03 have tested the predictions of their model against
observed color magnitude diagrams (CMDs) and integrated colors of star
clusters. Overall, their model provides excellent fits to star
clusters of different ages and metallicities in several photometric
bands.  They also compare the predictions of their model with observed
galaxy spectra in the Early Data Release of the Sloan Digital Sky
Survey (SDSS). Comparison of the strengths of various absorption
features and Lick indices between the SDSS galaxies, the BC03 models,
and other models shows a broad consistency between model and
observations.

The spectrum of a star forming galaxy, when observed over cosmological
distances, is modified significantly by two effects: (1) intergalactic
attenuation along the line of sight from the galaxy to the observer
and (2) dust attenuation in the emitting galaxy. In order to obtain a
realistic model of the observed galaxy spectrum, it is necessary to
account for both these effects.

\begin{figure*}
\plotone{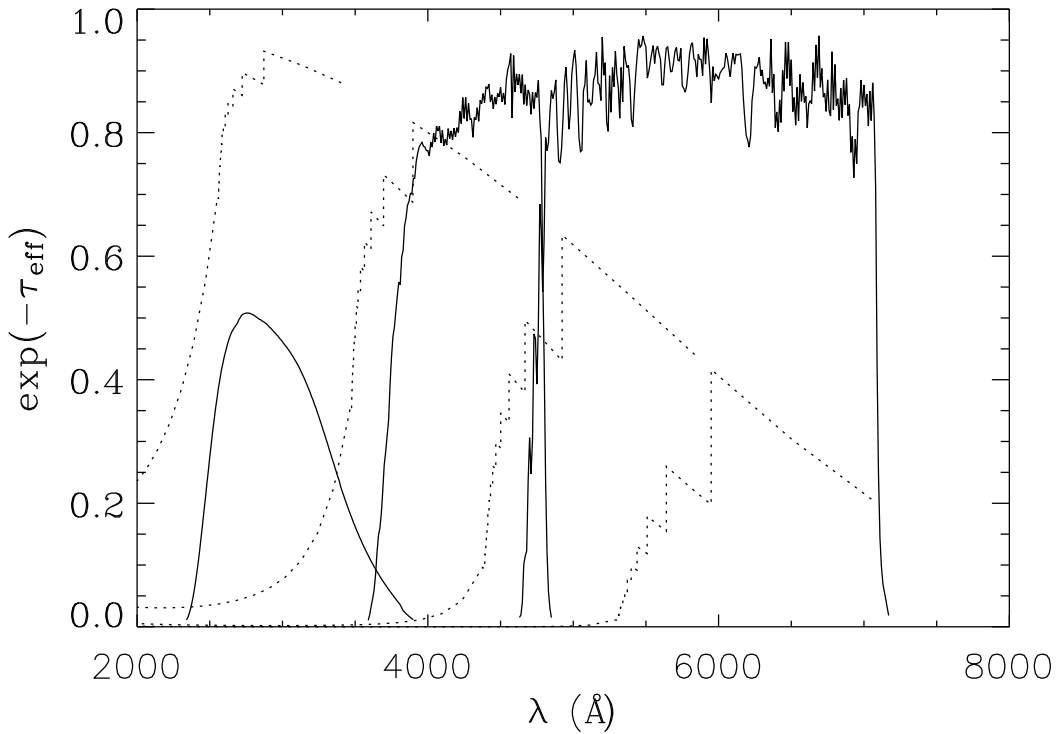} 
\caption{Mean transmission spectrum for a source at
$z$=1.8,2.8,3.8,4.8 (left to right, dotted lines) computed according
to the model of Madau (1995). The characteristic staircase profile is
due to continuum blanketing from the Lyman series. Also plotted are
the response functions of our $U$,$B$,$V$ filters (left to right,
solid lines).}\label{fig:transcurve}
\end{figure*}

\subsection{Intergalactic attenuation along the line of sight}\label{subsect:opacity}

The model of Madau (1995) for the propagation of UV radiation through
a clumpy universe is the standard work for determining intergalactic
attenuation. This model for the \HI opacity of the universe as a
function of redshift includes scattering in resonant lines of the
Lyman series (Ly$\alpha$, Ly$\beta$, Ly$\gamma$ and Ly$\delta$) and Lyman
continuum absorption. Recently, Meiksin (2006) has proposed a model
based on more current estimates of the properties of the intergalactic
medium (IGM). The results of numerical simulations were used to
estimate the contributions to resonant scattering from the
higher-order Lyman transitions. Differences of 0.5-1 mag from the
previous estimate of Madau are found. In this work, we have chosen to
use the standard work of Madau (1995) to enable us to make comparisons
with previous work that have almost without exception used this \HI
attenuation model.

The amount of flux attenuation can be computed as a function of
wavelength at each redshift of interest. We show in \fig{transcurve} a
plot of the transmitted power fraction as function of wavelength and
redshift. The characteristic staircase profile is due to continuum
blanketing by the Lyman series. At $z > 1.8$, the flux in the $U$-band
is significantly lowered by the increased \HI opacity. At higher
redshifts, flux in progressively redder bands is affected.

\subsection{Attenuation by dust in the emitting galaxy}\label{subsect:dustattenuation}

We use the simple but realistic prescription of Charlot \& Fall (2000)
to account for the attenuation of light by dust. In this prescription,
the attenuation of starlight by dust is accounted for by inserting a
factor $\exp[-\hat{\tau}_\lambda(t')]$ in the integrand on the
right-hand side of equation~(\ref{convol}) , where
$\hat{\tau}_\lambda(t')$ is the `effective absorption' curve
describing the attenuation of photons emitted in all directions by
stars of age $t'$ in a galaxy. This is given by the simple formula

\begin{eqnarray}
\hat{\tau}_\lambda(t')=\cases{ \phantom{\mu} 
 \hat{\tau}_V\left(\lambda/{5500\,\rmn{\AA}}\right)^{-0.7} \,,&for
 $t'\leq 10^7$ yr,\cr
 {{\mu\hat{\tau}_V}}\left(\lambda/{5500\,\rmn{\AA}}\right)^{-0.7}
 \,,&for $t'>10^7$ yr,\cr}
\label{taueff}
\end{eqnarray}
where $\hat{\tau}_V$ is the total effective $V$-band optical depth
seen by young stars. The characteristic age $10^7\,$yr corresponds to
the typical lifetime of a giant molecular cloud. The adjustable
parameter $\mu$ defines the fraction of the total dust absorption
optical depth of the galaxy contributed by the diffuse interstellar
medium ($\mu \approx 1/3$ on average, with substantial scatter).

\subsection{Our simulations}

The main aim in our simulations is to exploit the combined effect of
the intrinsic Lyman edge in galaxy spectra and the opacity of
intergalactic \HI gas to separate relatively unobscured star-forming
galaxies from those that are older and/or dustier, in a color-color
diagram of available broad-band colors.  Such a separation in
color-color space is most effectively achieved by plotting the $U-B$
color against the $B-V$ or $B-I$ color. In practice, the choice of
$B-V$ or $B-I$ does not have a significant impact on the efficiency or
completeness of the separation. Following C00, we chose the $B-V$
color as the redder color in our simulations.

\begin{deluxetable*}{ll}
\tablewidth{0pt}
\tabletypesize{\small}
\tablecaption{Parameter grid used for computing synthetic spectra.}
\tablehead{\colhead{}& \colhead{}}
\startdata
Ages& 0.001, 0.0025, 0.005, 0.0076, 0.01, 0.025, 0.05, 0.1, 0.5, 1.0, 5.0, 10.0 Gyr\\
Exponential star formation timescales& 0.01, 0.1, 1.0, 5.0, 10.0, 30 Gyr\\
Constant star formation durations& 0.01 Gyr\\
Metallicities& 0.2, 0.4, 1.0 $Z_{\odot}$\\
$\hat{\tau}_\lambda(t')$& 0.01,0.51,1.01,1.51,2.01,2.51\\
$\mu$&0.02, 0.22, 0.42, 0.62, 0.82\\
Redshifts& 0.2-7.0 Stepsize: 0.2\\
\enddata
\label{tab:parametergrid}
\end{deluxetable*}

We wish to compute colors for galaxies populating a grid of ages,
metallicities, dust obscuration and star formation histories at
various redshifts (see \tab{parametergrid} for a summary of parameter
grid values used in our simulations). As a first step, we need to
study how the choice of IMF (Salpeter or Chabrier) affects the
distribution of galaxies in the color color space. The Chabrier IMF is
identical to the Salpeter IMF for stars that are more massive than 1
$M_\odot$. For lower mass stars and substellar populations such as
brown dwarfs the two IMFs are significantly different. However, the
rest-frame UV emission from galaxies at high redshift that is probed
by the colors used in our simulations, receives a nearly negligible
contribution from low mass stars and brown dwarfs. Thus, for our
simulations, choosing either IMF should produce nearly
indistinguishable color-color plots. We verified this by comparing
color color plots for both IMFs in the relevant region of color-color
space. For consistency with previous work, we chose to use the
standard Salpeter IMF in our models. There is no prior on which points
on the grid are valid, except the obvious logical requirement that the
galaxy should be younger than the age of the universe at that
redshift. The grid thus samples the plausible range of galaxy
properties, without weighting the results with our knowledge of the
actual distribution of galaxy color that may be derived from either
observations or theoretical models.

We compute the spectral energy distribution (SED) of each galaxy by
modeling it as a composite stellar population that formed either in a
constant burst of star formation lasting $10^7$ years or in an
exponentially decaying burst with timescales ranging from $10^6$ years
to $3 \times 10^{10}$ years. The SEDs are computed at a range of ages
(defined as the time since the onset of star formation) and with a
range of metallicities. Note that all stars in a given galaxy have
exactly the same metallicity in the BC03 model. Due to this, the
change in integrated colors due to metallicity gradients within a
galaxy are not modeled. The SEDs are corrected for dust attenuation
(\subsect{dustattenuation}), appropriately redshifted and then flux
corrected for cosmic opacity along the line of sight
(\subsect{opacity}).  Finally the SEDs in the observer frame are
folded through the filter response functions and the broad-band colors
in the $U-B$ and $B-V$ filter are obtained.

\begin{figure*}
\plotone{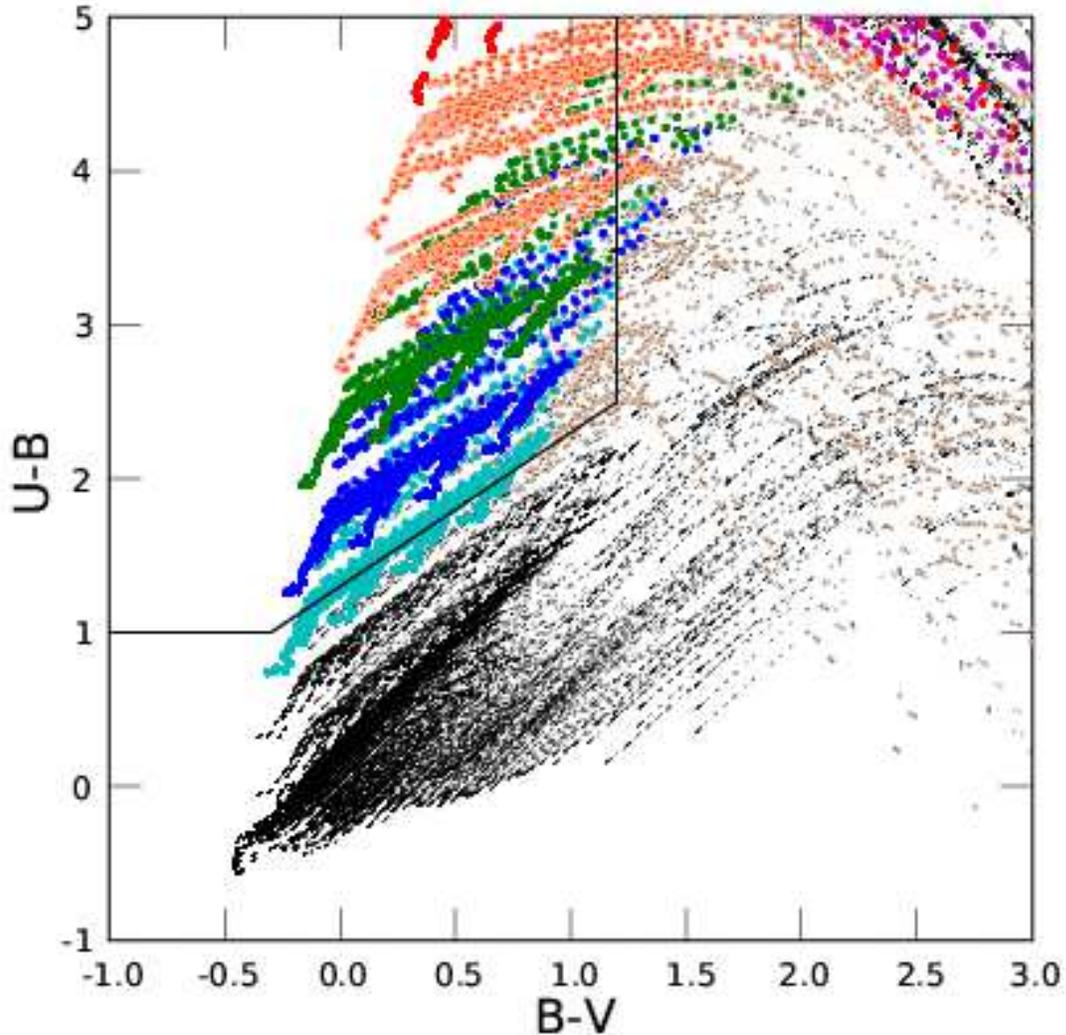} 
\caption{$U-B$ versus $B-V$ for our simulated galaxies. A total of 264
600 synthetic spectra of galaxies representing a wide range of ages,
star formation histories, metallicities, dust properties and redshifts
were folded through our combination of WFPC2 and ACS filters. Galaxies
shown in colors other than black represent young star-forming galaxies
with relatively low level of dust obscuration in the redshift range
$2.0 < z < 3.5$. The colors cyan, blue, green, orange, red and magenta
show galaxies with progressively higher redshift in bins of 0.25 per
color. Galaxies in the same redshift range that are old and/or
severely dust attenuated are shown as gray dots. Black dots represent
galaxies outside the redshift range. Our color selection criterion is
shown by the polygon. Galaxies within the polygon are the U-band
dropout candidate galaxies with $2.0 < z < 3.5$. The selection
criteria are $U - B > 1.0$, $ U - B > B - V + 1.3$, and $B - V <
1.2$. }\label{fig:simulatedcolors}
\end{figure*}

\subsection{Colors of relatively unobscured star forming galaxies at
$2.0 \lesssim z \lesssim 3.5$}\label{subsect:colors}

We show in \fig{simulatedcolors} the color-color diagram that we
obtained for our simulation of galaxies using the parameter grid
described above. Colored symbols (cyan through magenta) identify
galaxies within the desired redshift range $2 \lesssim z \lesssim 3.5$
that are both starforming (age $<0.1$ Gyr) and relatively unobscured
by dust ($\hat{\tau}_\lambda(t') < 2.0$).  Brown symbols identify old
and/or attenuated galaxies in the same redshift range; such galaxies
are typically much fainter in restframe UV (observed $ B $) and are
unlikely to constitute a large fraction of our sample.  Black dots
identify galaxies outside the target redshift range.

We find that relatively unobscured star forming galaxies at $2\lesssim
z \lesssim 3.5$ are effectively selected in color-color space by the
following color criteria: (1) $U - B > 1.0$ (2) $ U - B > B - V + 1.3
$ and (3) $B - V < 1.2$. The minimum ``redness'' threshold in $U-B$
color (horizontal line) is motivated by the desire to exclude objects
with $z < 2$. For a given metallicity and dust attenuation, galaxies
with $2\lesssim z \lesssim 3.5$ become redder in $U - B$ at
progressively higher redshifts while remaining relatively unaffected
in their blue $B - V$ color. As the redshift approaches $z \lesssim
3.5$, the Lyman $\alpha$ forest and eventually the Lyman limit begins
to progressively intrude into the $B$ band causing the $B-V$ color to
become redder. This begins to move galaxies rightward and out of the
selection box. Increasing metallicity and dust obscuration both tend
to make objects redder; the extent of the effect on each color is
strongly dependent on the age of population and the redshift.

By design, these selection criteria are aimed at identifying galaxies
with strong UV flux.  Therefore our sample will exclude galaxies with a
low star formation rate or with strong reddening.  The galaxies to the
right of our $B-V$ cutoff in \fig{simulatedcolors} are relatively dusty
($\hat{\tau}_\lambda(t') \gtrsim 1.0$) and are forming stars at a
relatively low rate.  Such moderately reddened galaxies with modest
star-formation rates have been identified and studied spectroscopically
at brighter magnitudes at somewhat lower redshift (Daddi et al.  2004;
Reddy et al.  2005), but their luminosity function has not been probed
directly to the faint levels we are investigating here.  Similarly,
recent ground-based studies (e.g., Le F{\`e}vre et al.  2005) find a
non-negligible number of spectroscopically confirmed high redshift
galaxies below the selection box (bluer $U-B$ color).  Again, these
objects are significantly brighter ($I < 24$) than the objects that we
aim to study in this work.  Significantly expanding the selection box in
either direction would include many more contaminating objects outside
our target redshift range; therefore we prefer to maintain a
conservative selection criterion that keeps contamination low, while
including the majority of strongly star-forming, relatively unobscured
galaxies.  This choice is consistent with that of other
dropout studies, although our color limits are slightly different
from C00 due to the updated stellar population models and to
the fact that the ACS $B$ (F435W) filter we use is shifted blueward 
of the WFPC2 $B$ (F450W) filter.

\begin{figure*}
\plotone{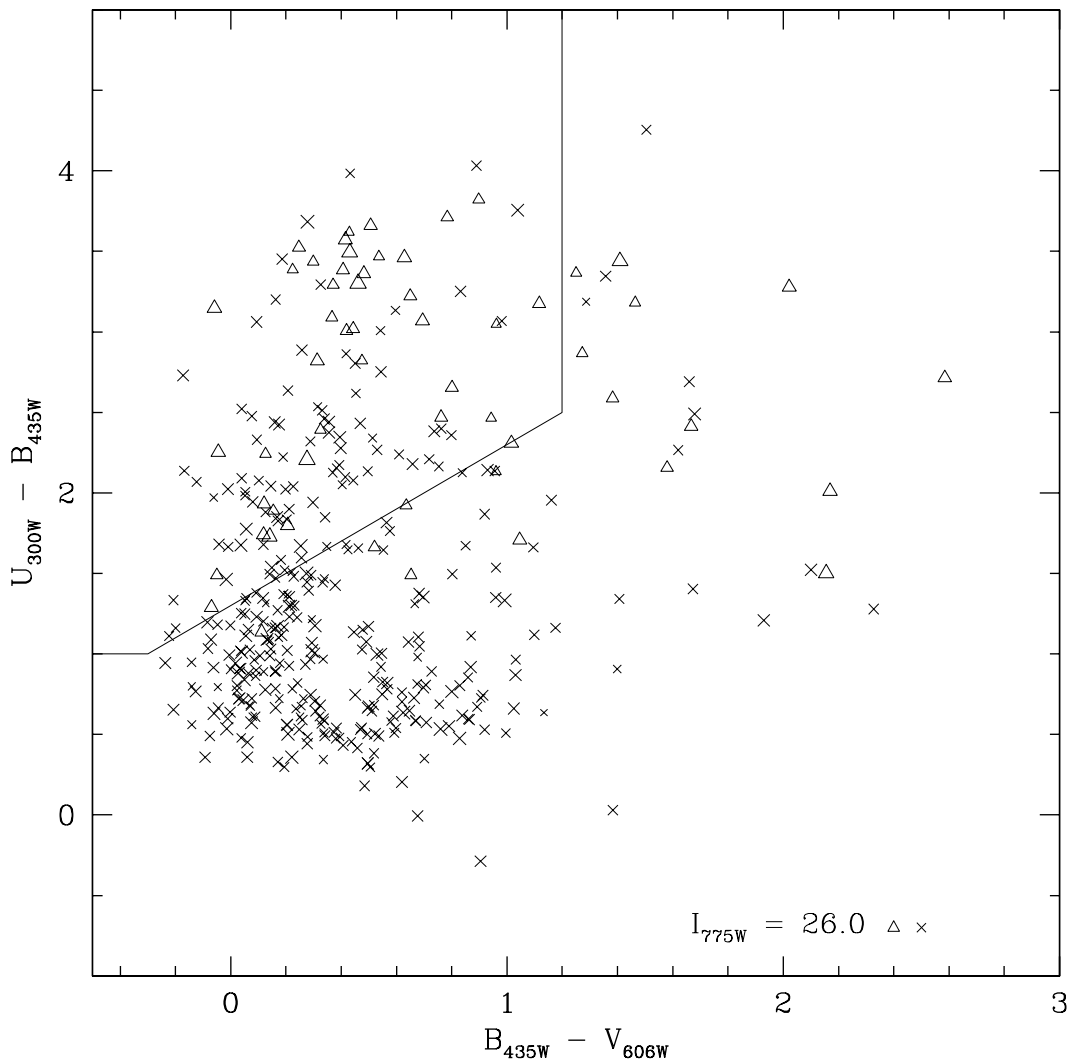} 
\caption{Color-plot indicating the candidate $U$-band dropouts
(enclosed by the polygon) in our data. Crosses indicate detections;
triangles indicate non-detections in the $U$-band and are placed at
the 1$\sigma$ lower limit of the $U-B$ color. The size of the symbols
scales with magnitude in ACS/F775W.}\label{fig:ubbv}
\end{figure*}

\subsection{High redshift galaxies in the HUDF parallels: The sample}

Having identified the color selection criteria above, we are now in a
position to identify $U$ dropout galaxies in our data. To obtain
photometry for our images, we used Sextractor version 2.3.2 (Bertin \&
Arnouts 1996) in dual image mode. We used the GOODS/ACS $i$ band image
for source detection and the WFPC2 $U$-band and GOODS/ACS $B$ and $V$
band images for photometry. The region of the GOODS data that fell
outside the footprint of the $U$ band image was set to zero. 5.67
sq. arcmin (87\%) of the 6.51 sq. arcmin footprint of our $U$-band
image overlaps with the GOODS coverage area (see \fig{final}).

The main advantage of dual mode photometry is that there is no need to
make a subjective decision to match corresponding sources detected in
the various bands. Setting up such a correspondence is particularly
difficult in situations where a single object in one filter gets
detected as multiple objects in another filter, if there are multiple
peaks in the light distribution. Such a situation is common in the $U$
band which selectively probes the regions of ongoing star formation
within a galaxy. Detection was carried out with a {\tt
DETECT\_MINAREA} of 12 pixels and a {\tt DETECT\_THRESH} of 1.5. The
detections were filtered with a Gaussian kernel with FWHM 2.5 pixels
and cleaned with {\tt CLEAN\_PARAM}=1.0. Object deblending was carried
out with {\tt DEBLEND\_MINCONT}=0.03

The sample is first truncated at magnitude limits sufficiently bright
that Lyman breaks of the expected amplitude could be measured
reliably.  For our data, this corresponds to $B < 27.3 $. We find that
all objects that are brighter than this limit in the $B$ band are also
detected in the $i$ band image. We show in \fig{ubbv}, the color-color
plot highlighting the candidate $U$-band dropouts.  Objects with
measured magnitudes in all three bands are indicated by crosses.
Objects with measured $ U $ band flux (as indicated by the Sextractor
{\tt MAG\_AUTO} magnitude) fainter than the 1$\sigma$ limit computed
from the inverse variance map at that location are considered
undetected, and are marked by a triangle at the $U-B$ color
corresponding to the 1$\sigma$ magnitude limit. The 1$\sigma$ limit
depends on the size of the object. Additionally, as the exposure time
is not uniform across the image, it also varies with position.  The
size of each symbol is proportional to the $i$ band magnitude of the
object.

The polygonal selection region which identifies $ U $ band dropouts,
as defined in \subsect{colors}, contains 136 objects, each of which
has been visually inspected. Five objects, all relatively bright ($ i
< 22 $), appear to be stars; 6 more objects are otherwise spurious,
either due to instrumental artifacts (diffraction spikes, image edges)
or resolved segments of nearby bright galaxies.  These 11 objects are
not plotted in \fig{ubbv} and have been excluded from the subsequent
analysis. The remaining clean sample of $U$ dropouts consists of 125
objects; their coordinates and magnitudes are listed in
\tab{dropoutlist}.

\begin{figure*}
\epsscale{.8}
\plotone{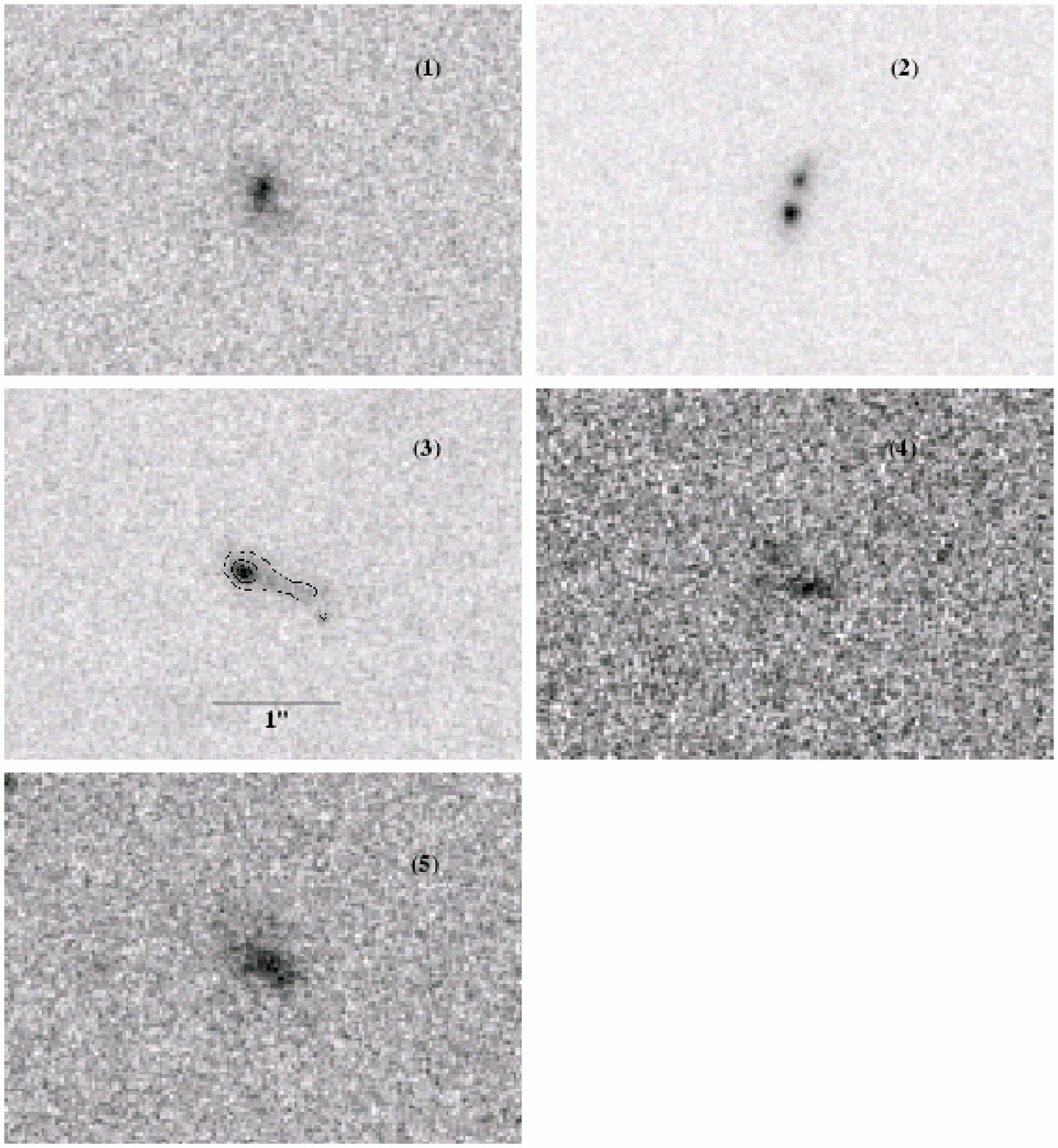} 
\epsscale{1}
\caption{Representative dropout in each of the 5 classes into which we
classify objects. The classes are (1) clumps, (2) double or multiple
clumps, (3) clumps with tails, i.e.  tadpoles, (4) objects which are
not compact but extended and fuzzy and (5) fuzzy extended objects but
with a more organized disk-like structure. }\label{fig:class}
\end{figure*}

\begin{figure*}
\plotone{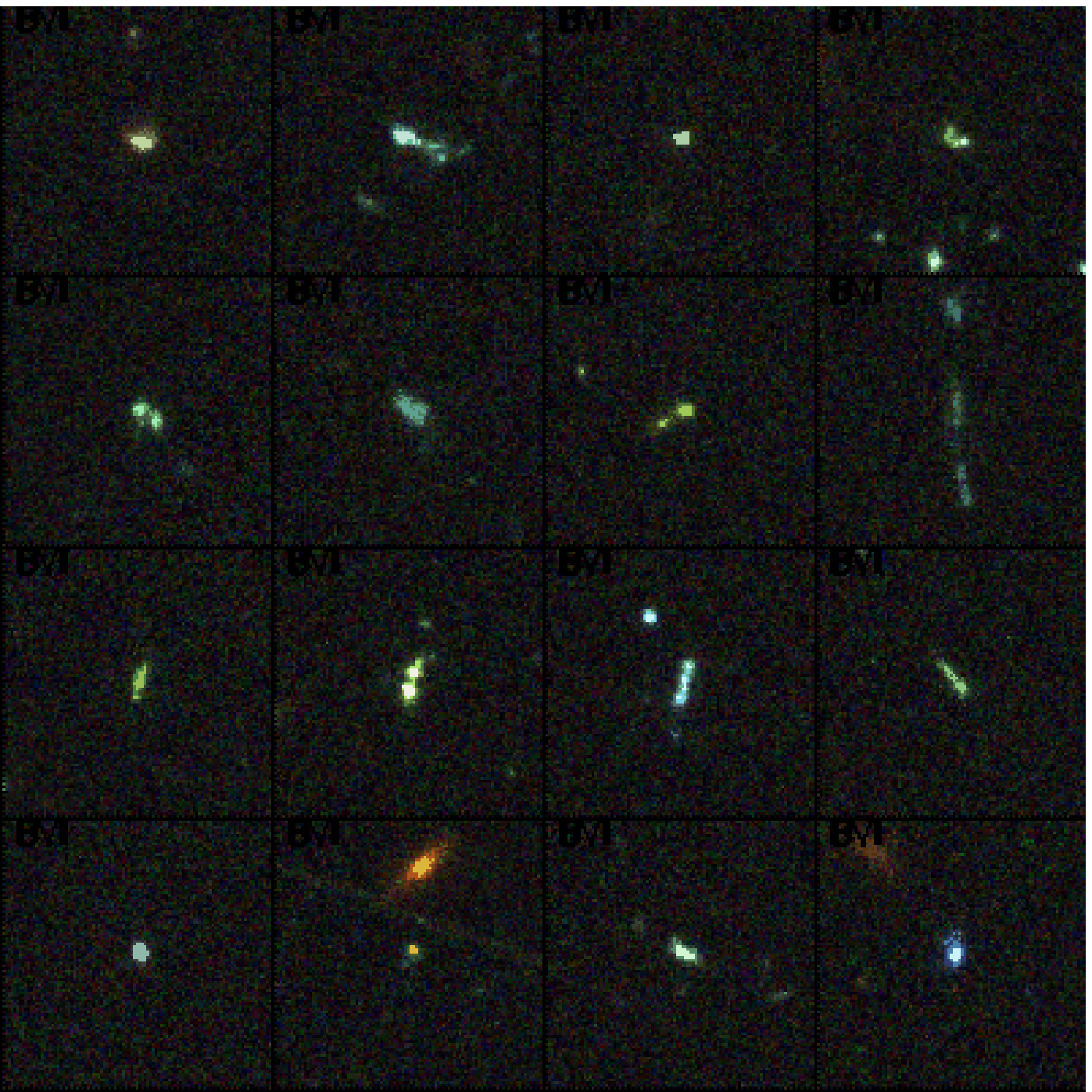}
\caption{BVi color composite of 16 bright $U$ band dropouts in our
sample. Each cutout is 5x5 arcsec in size. The dropout galaxy is
located at the center of each panel.}\label{fig:morphology}
\end{figure*}

\section{Dropout Morphology}

At redshifts $z\sim 3$, the drizzled ACS images with a resampled pixel
scale of 0.03 arcsec/pixel probe regions as small as 0.23 kpc/pixel
for our chosen cosmology.  The fine resolution provides sufficient
resolution elements to enable at least a crude study of their
morphology. The drizzled GOODS/ACS images have the advantage of a
pixel scale that is a factor of two finer than the drizzled HDF
images. However, the depth reached by the GOODS survey is about a
magnitude shallower than the corresponding HDF data. Morphological
details can thus be studied with better resolution but poorer
signal-to-noise, relative to the HDF images.

We have visually inspected the entire sample of $U$-dropouts in the
original GOODS/ACS $z$-band images and classified their morphology
using an empirical classification scheme.  Objects that appear to be
single and compact are classified as clumps (1), double or multiple
clumps (chains) are classified as (2), clumps with tails, i.e.
tadpoles as (3).  Objects which are not compact but extended and fuzzy
are classified as (4), and fuzzy extended objects with a more
organized disk-like structure are classified as (5).  \fig{class} illustrates
typical objects of each class. We list in \tab{histmorph} the relative
fraction of each class.  For our sample, the fraction of single clumps
is similar to the fraction of fuzzy objects ($\sim$ 35\%), however if
double clumps and tadpoles are included in the clump class (1), this
fraction increases to $\sim$50\%. The morphological classification for
each object and short remarks on its appearance are listed in
\tab{dropoutlist}. We also show in \fig{morphology} $BVi$ color
composites of 16 typical, bright dropouts in our sample.

Since the original deep fields it has been known that compact/clump
objects dominate at faint magnitudes (Abraham et al. 1996, van den
Bergh 2002). Recently, this result has been confirmed by the
morphological studies of Ultra Deep Field where most of the galaxies
are found to be clumps (Elmegreen et al. 2005a, 2005b). An important
question is whether such clumpy structures are signatures of
merging/accretion or a merely band-shifting effect. The latter is
probably not the case since these clumps have no counterparts in the
local Universe (Elmegreen et al.  2005a, 2005b, Windhorst et
al. 2002). Moreover, as shown in Papovich et al. (2005) for galaxies
at $1.9<z<3$, there is very little morphological transformation
between wavelengths using WFPC2 F606W image and NICMOS F160W which
spans a long wavelength baseline (rest-frame far-UV to the rest-frame
B band).

We have also checked whether the morphological counterparts of
$U$-dropouts are seen in the lower redshift ($z <2$) sample (de Mello
et al. 2006) for which we have photometric redshifts from GOODS
(Dahlen et al. 2005).  We found that $U$-dropout objects show
morphological similarities to ones classified in de Mello et
al. (2006) as starbursts\footnote{Spectral type obtained from the
template fitting in the photometric redshift technique (Dahlen et
al. 2005).} which are classified as `compact, peculiar or low surface
brightness (LSB)'. The objects classified as `compacts' are similar to
class (1). Objects classified as (2) and (3) would fit into the de
Mello et al. `peculiar' class, and objects classified as class (4)
(fuzzy) correspond to the LSB class. However, as expected, the
$U$-dropout sample lacks the classical Hubble types, such as
elliptical/spheroid and disks which are seen at lower redshifts
($z<0.8$). The fraction of LSBs and compact+peculiar objects in the
sample of de Mello et al.  (2006) which are at $0.8<z<1.2$ is 27\% and
52\%, respectively. Interestingly, the corresponding fractions of
$U$-dropouts are very similar, 38\% and 51\% (extended fuzzy and
clump+doubles+tadpoles) indicating that about 50\% of the star
formation takes place in clumps over an extended redshift range from
$0.8 \lesssim z \lesssim 3.5$.

\section{Dropout statistics}

It is interesting to compare the statistical properties of the
dropouts selected here with those in the two HDFs.  The selection
procedures are very similar, although slight differences can arise
from the different filter set available and from the different depth
of the observations---those used here are significantly deeper in $U$,
but shallower in the other bands.  With those caveats in mind, we
compute some properties of interest for our sample.

\subsection{Number density}

We have identified 125 $U$ band dropouts in the 5.67 square arcmin of
sky over which we have multiwavelength data available.  Of these, 80
fall within the full depth coverage of our F300W data which covers
4.02 square arcmin. The surface density of dropouts at the full depth
of our study is thus 19.9 sources per square arcmin, similar ---within
the statistical uncertainties--- to the 17.5 sources per square arcmin
found in the HDF-S data (C00). Relative to the HDF-N value of 16.0
sources per square arcmin our measured surface density is higher with
a marginal significance. Both the HDF measurements were obtained using
the Madau et al. (1996) color selection criteria and only included
sources brighter than $B_{450}=26.79$. Our sample uses a different
filter combination with its own color selection criterion and all
sources brighter than $B=27.3$.

Using only the dropouts in the region with full depth and assuming
that these dropouts uniformly probe the redshift range $2 \lesssim z
\lesssim 3.5$, we obtain a comoving galaxy density of $4.0 \times
10^{-3}$ Mpc$^{-3}$ for the concordance cosmology we adopt,
corresponding to $5.7 \times 10^{-3}$ Mpc$^{-3}$ with the cosmology
adopted by Madau {\etal} (1996; $q_0=0.5, h_{100}=0.5$, no
cosmological constant).  For comparison, they report a density of $4.2
\times 10^{-3}$ Mpc$^{-3}$ from the HDF-N.  Our somewhat higher galaxy
density may be due in part to the increased depth of our sample,
although cosmic variance can also play an important role on such small
areas.

\subsection{Star formation rate density evolution}

A key measure is the evolution of the cosmic star formation rate (SFR)
density as a function of lookback time (or redshift) (e.g., Madau
{\etal} 1996). The star formation rate may be measured using carefully
calibrated proxies such as the luminosity in X-ray (Ranalli et
al. (2003), radio (Condon et al. 1992), far-infrared (Kennicutt 1998),
narrow-band line emission (Moustakas et al. 2006) and the rest frame
UV continuum (Madau et al. 1998). When such indicators are used to
compute the SFR for the {\it same} sample, they are consistent to
within a factor of two (Daddi et al. 2004).

Even within various samples identified using the dropout technique,
significant discrepancies may exist. These may be caused by cosmic
variance, differences in stellar population synthesis models or in the
treatment of dust attenuation. Different assumptions on the shape of
luminosity function and the differing photometric uncertainties in
observations of different depths can also introduce systematics into
the measurement. Of these, the uncertainties in the dust obscuration
correction are of the greatest concern in estimating the SFR from the rest frame UV continuum. Recently, Heckman et
al. (2005) have obtained GALEX observations of compact UV luminous
galaxies in the local universe. They find that these galaxies, which
are chosen to have rest-frame UV luminosities similar to those of
$z\sim3$ star forming galaxies also match them in stellar mass,
velocity dispersions and gas phase metallicity. Like their distant
counterparts, these galaxies are forming stars at a rate high enough
to build their present day stellar mass in 1-2 Gyr. For these local
galaxies, dust attenuation is relatively modest -- 0.5-2
magnitudes. If the high redshift galaxies exhibit similar dust
properties, the effect of uncertainty in dust attenuation on the
computed star-formation rate may be less than previously thought
(Vijh, Witt \& Gordon 2003).

\begin{figure*}
\epsscale{.8}
\plotone{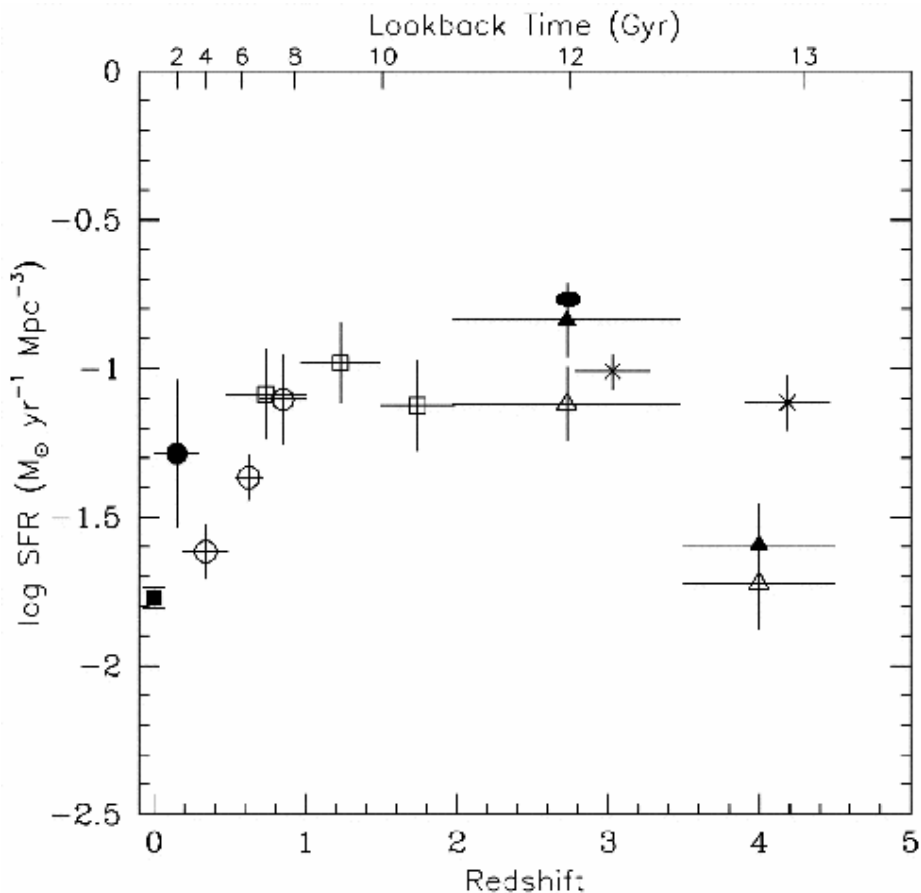} 
\epsscale{1}
\caption{SFR density evolution as a function of redshift (compare to
Figure 11 of C00). The SFR is estimated from the UV luminosity
density. The $z > 2$ points are from Lyman-break objects in the HDF-N
(open triangles), in the HDF-S (filled triangles) from C00, and in the
Steidel {\etal} (1999) ground-based survey ($\times$ symbols). The
black ellipse indicates the SFR density measured in this work. For
clarity, error bars are not plotted for our datapoint. Errors along
the redshift axis are identical to the HDF points. The statistical error on the SFR density axis is about 0.09. Distances and volumes
are computed using the cosmological parameters $h,\Omega_{\rm m},
\Omega_\Lambda, \Omega_{\rm tot} = 0.65, 0.3,0.7,1.0.$. At
lower-redshifts, only results that are directly comparable to the high
redshift data are plotted. The open squares are from HDF photometric
redshifts by Connolly et al (1997), the open circles are from Lilly
{\etal} (1996), the solid square is from the H$\alpha$ survey of
Gallego {\etal} (1995), and the solid circle from Sullivan {\etal}
(2000).}\label{fig:sfr}
\end{figure*}

In spite of these uncertainties, general trends in the evolution of the
SFR density as computed from the UV continuum luminosity have become
apparent over the last decade. We now know with a high level of
confidence that the SFR density steadily increases from the local
Universe out to a redshift of $z \sim 3$ and then either shows a fall
off or remains steady at higher redshifts (e.g., Lilly et al.\ 1996;
Madau et al.\ 1996; Dickinson\ 1998; Steidel et al.\ 2003; Bouwens et al.\ 2004).  We adopt
here the formulation of C00, which in turn follows the prescriptions
of Steidel et al.\ (1999).  For ease of comparison with Figure 11 of
C00, we use their cosmological parameters
$h,\Omega_m,\Omega_{\Lambda},\Omega_{\rm tot} = 0.65,0.3,0.7,1$ and
define a fiducial magnitude $R_{606+775}$ for each candidate, which is
the average of the $V$ and $i$ magnitudes. We use this magnitude to
compute $m_{\rm AB,tot}$ which is the total integrated flux of the
Lyman-break candidates. To avoid variable levels of incompleteness, we
only use dropouts in the full depth region of our image for computing
the integrated flux. Then, the star-formation rate density at a mean
redshift $z$ between two redshifts $z_1,z_2$ is given by

\begin{eqnarray}
   \log ({{SFR} \over {M_\odot \rm yr^{-1} Mpc^{-3}}} )  = 
		2 \log {{D_L(z)} \over {\rm cm}} - 0.4 m_{\rm AB,tot }\cr
		- \log {{\Delta V(z_1,z_2)} \over {\rm Mpc^3}} - \log A\cr
		+ \log \delta L - \log(1+z)
		- 34.516,
 \end{eqnarray}
where the redshifts $z,z1,z2$ are 2.75,2.00,3.50 respectively, $D_L$
is the luminosity distance, $\Delta V$ is the total volume between the
two redshifts, and $A$ is the area of the field of view in square
arcseconds.  The contribution of galaxies fainter than our detection
threshold is included via the correction factor $\delta L$, which is
obtained by integrating the assumed luminosity function (Steidel et al.\ 1999) from our survey limit down to $0.1 L^*$ (Formula 9,
C00). The above formula includes a uniform extinction correction
factor of 4.7 (Steidel et al.\ 1999) to the measured UV luminosity
densities. This correction factor is appropriate for a typical
extinction $E(B-V)=0.15$ and the Calzetti (1997) reddening law.

We measure an integrated star-formation rate density of $0.18\  {M_\odot
\rm yr^{-1} Mpc^{-3}}$. This value is only marginally higher than that
measured for the HDF-S, which in turn is higher than that measured for
the HDF-N. Both these values were measured by the authors of C00,
using identical source detection and photometry procedures with data
of similar depth obtained using the same set of filters and with the
same color selection criteria. We show in \fig{sfr} the evolution of
star formation rate density with redshift, comparing our measurement
with other measurements from the literature obtained using a similar technique.

Measurements of the statistical properties of any clustered population
are plagued by uncertainty due to cosmic variance, the field to field
variation due to large scale structure (Somerville et al. 2004). At
high redshifts, especially with the small few arcmin sized field of
view of HST, the volume sampled between $2.0 \lesssim z \lesssim 3.5$
are $\sim10^4$ Mpc$^{3}$, small enough that cosmic variance is a
significant source of uncertainty. The variance of number counts $N$ over an area $A$ on the sky, in the presence  of a non-zero angular auto-correlation function is given by (Peebles, 1980):

\begin{equation}
\sigma^2(N)= NA + n^2 \int_{A_1} dA_1 \int_{A_2} dA_2 w(\theta)
\label{variance}
\end{equation}
where $n$ is the average density of objects and thus $nA$ is the
expectation value of $N$, and $w(\theta)$ is the two-point angular
auto-correlation function between points $A_1$ and $A_2$. For our full
depth area of 4.02 sq. arcmin the second term in equation
\ref{variance} can be approximated as $1.8n^2A^2A_w$ where $A_w$ is
the two point angular correlation of the sample at a scale of 1
arcmin. This assumes that the angular correlation function obeys a
power law with exponent $\beta=-1.2$ as measured by Giavalisco \&
Dickinson (2001) for a combined sample of $U$ dropouts from the two
HDFs. We estimate the value of $A_w$ at 1 arcmin from the $1\sigma$
upper limit on their measured angular correlation function for their
combined HDF sample as $A_w=0.02$. The total expected variance is
thus:

\begin{equation}
\sigma^2(N)= 80 + 1.8 \times 0.02 \times 80^2 = 310.4
\label{variance}
\end{equation}
or approximately 3.9 times larger than the Poisson contribution
alone. The expected r.m.s. fluctuation in number counts is about 17.6
galaxies over the full depth area, compared with 8.9 for the Poisson
case. This corresponds to an uncertainty of 0.09 in the logarithm of
the star formation rate density, and may be treated as the nominal
statistical uncertainty on our measurement. It should be noted that
this error bar only accounts for statistical uncertainties; systematic
offsets e.g. due to incorrect dust corrections may be significantly
larger.

An effective way to reduce the uncertainty caused by clustered
populations is to obtain samples of high redshift galaxies along
independent lines of sight. Our measurement of the SFR density in the
GOODS-S area provides such an independent measurement, and is in fact
only the third such measurement obtained exclusively from space-based
data at $z \sim 3$.  Within the error bars, our measurement of SFR
density is consistent with the value measured for HDF-S, indicating
that the range of variation in SFR density is probably well sampled by
the two HDFs.

In the future, the second WFPC2/F300W parallels, if combined with
additional new data in the $B,V$ filter can provide a fourth
space-based measurement of the SFR density at $z \sim 3$, further
reducing the impact of cosmic variance.
  
\acknowledgements

We thank the GOODS and HUDF teams for the magnificent public data that
made this work possible. We are grateful to the referee G. Zamorani
for insightful comments and suggestions that helped improve the
content and presentation of this paper. We thank Harry Ferguson and
Uma Vijh for fruitful discussions on different aspects of this paper.
Support for program AR 9540 was provided by NASA through a grant from
the Space Telescope Science Institute, which is operated by the
Association of Universities for Research in Astronomy, Inc., under
NASA contract NAS 5-26555.

\clearpage

\pagestyle{empty}
\LongTables
\begin{deluxetable}{l l r l l l l l}
\tabletypesize{\small}
\tablewidth{0pt}
\tablecaption{$U$ band dropouts.}
\tablehead{\colhead{$\alpha_{\rm J2000}$ (degrees)}& \colhead{$\delta_{\rm J2000}$ (degrees)}&\colhead{$U$}&\colhead{$B$}&\colhead{$V$}&\colhead{$i$}&\colhead{Class}&\colhead{Remarks}}
\startdata
       53.137048&      -27.712829&$$      27.54&      24.48&      24.39&      24.30& 3&    tadpole\\					       
       53.145949&      -27.710505&$$      27.08&      25.13&      25.05&      24.97& 4&   fuzzy\\						       
       53.164155&      -27.709900&$$      29.11&      25.86&      25.03&      24.74& 5&   faint clumpy disk\\				       
       53.130524&      -27.709571&$>$      28.18&      25.36&      25.05&     24.40& 5&   disk/clumpy or double clump\\			       
       53.157444&      -27.709015&$$      29.17&      25.13&      24.24&      23.99& 2&   double clump\\					       
       53.171374&      -27.709446&$$      28.82&      26.44&      26.09&      25.93& 1&   faint clump\\					       
       53.156550&      -27.708817&$$      28.48&      26.96&      26.77&      26.80& 1&   faint clump\\					       
       53.166266&      -27.708565&$$      28.08&      26.23&      26.06&      26.08& 1&   clump\\						       
       53.155762&      -27.708305&$$      29.53&      26.47&      25.48&      25.12& 5&   clumps in disk\\				       
       53.166611&      -27.708280&$$      28.57&      26.53&      26.38&      25.89& 4&   faint fuzzy\\					       
       53.160557&      -27.707669&$$      27.35&      25.68&      25.33&      24.77& 2&   multiple clump (maybe on faint disk)\\		       
       53.130624&      -27.707912&$>$      28.86&      27.12&      27.00&     26.39& 5&   faint clumpy disk/tail?\\			       
       53.150609&      -27.707880&$$      27.90&      25.67&      25.48&      25.47& 4&   faint fuzzy\\					       
       53.153684&      -27.707186&$$      27.34&      25.25&      24.83&      24.63& 4&   fuzzy/clump\\					       
       53.150374&      -27.706826&$$      27.59&      25.05&      24.74&      24.46& 5&   faint clumpy disk\\				       
       53.131003&      -27.706498&$$      26.94&      25.27&      25.23&      24.65& 1&   clump\\						       
       53.143096&      -27.705956&$$      29.69&      26.80&      26.54&      26.50& 5&   faint clumpy disk\\				       
       53.175579&      -27.705782&$$      27.35&      25.20&      24.82&      24.62& 2&   double clump - merging\\			       
       53.130086&      -27.705758&$>$      28.89&      26.69&      26.41&     26.31& 4&   faint fuzzy\\					       
       53.173130&      -27.705631&$$      28.91&      26.77&      26.28&      26.07& 4&   faint fuzzy\\					       
       53.182801&      -27.705272&$$      27.09&      24.57&      24.53&      24.55& 1&   clump edge on\\					       
       53.135928&      -27.705299&$$      29.00&      26.97&      26.98&      26.86& 4&   faint fuzzy\\					       
       53.181991&      -27.705029&$$      28.49&      26.90&      26.72&      26.76& 4&   fuzzy\\						       
       53.150622&      -27.704625&$$      27.38&      25.40&      25.35&      25.32& 1&   clump offcenter\\				       
       53.174787&      -27.704499&$>$      30.11&      26.65&      26.02&     26.09& 4&   faint fuzzy close to spiral\\			       
       53.148378&      -27.704298&$$      29.81&      26.06&      25.02&      24.72& 1&   clump (spheroid)\\				       
       53.165698&      -27.703950&$>$      29.90&      26.83&      26.14&     25.94& 4&   very faint fuzzy\\				       
       53.131169&      -27.703968&$$      27.86&      25.85&      25.80&      25.74& 1&   clump/edge on\\					       
       53.130501&      -27.703690&$$      28.37&      26.23&      26.40&      25.77& 5&fuzzy disk\\					       
       53.175991&      -27.703135&$>$      30.43&      26.91&      26.66&     26.67& 1&   faint clump\\					       
       53.149082&      -27.702952&$$      28.75&      26.02&      26.20&      26.24& 1&   clump/edge on\\					       
       53.141871&      -27.702771&$$      28.44&      27.11&      27.31&      27.01& 4&   fuzzy\\						       
       53.137756&      -27.701292&$>$      29.65&      26.26&      25.85&     25.71& 1&   faint clump offcenter on fuzzy tail (companions)\\     
       53.129540&      -27.701279&$>$      28.96&      27.23&      27.09&     26.74& 1&   clump next to peculiar shape disk (multiple clump?)\\  
       53.141437&      -27.701159&$$      28.99&      27.11&      26.99&      26.53& 6&   very faint\\					       
       53.138861&      -27.701083&$$      27.26&      25.67&      25.41&      25.01& 1&   clump\\						       
       53.178458&      -27.701173&$$      28.77&      27.23&      27.08&      26.33& 5&   faint disk edge-on\\				       
       53.138012&      -27.701123&$>$      29.91&      26.55&      26.07&     25.91& 1&   clump/close to other faint clumps\\		       
       53.161515&      -27.684979&$$      28.00&      26.06&      25.77&      25.59& 1&   clump/tail close to fuzzy\\			       
       53.161355&      -27.685116&$$      27.33&      25.65&      25.69&      25.86& 4&   fuzzy close to clump/tail\\			       
       53.145630&      -27.685261&$>$      27.88&      25.41&      24.65&     24.44& 1&   clump(spheroid)?\\				       
       53.165419&      -27.685627&$$      28.87&      26.59&      26.19&      26.07& 4&   fuzzy\\						       
       53.142972&      -27.685698&$$      26.93&      25.47&      25.49&      25.27& 5&   faint disk edge-on\\				       
       53.140820&      -27.685880&$>$      28.09&      26.60&      26.65&     26.47& 5&   fuzzy/disky\\					       
       53.143404&      -27.686231&$>$      28.72&      26.83&      26.68&     26.53& 2&   double clump\\					       
       53.147907&      -27.687704&$$      27.95&      26.27&      26.15&      26.26& 2&   big chain\\					       
       53.146281&      -27.688363&$>$      28.47&      26.68&      26.47&     26.33& 4&   fuzzy close to fuzzy objects\\			       
       53.147855&      -27.688591&$>$      28.37&      26.44&      26.32&     26.21& 2&   chain\\						       
       53.161346&      -27.688836&$$      28.73&      26.30&      25.83&      25.55& 4&   fuzzy/edge on\\					       
       53.154624&      -27.689418&$>$      30.31&      27.16&      27.22&     27.30& 1&   faint clump\\					       
       53.143049&      -27.689724&$>$      28.62&      26.38&      26.25&     26.21& 4&   fuzzy\\						       
       53.169292&      -27.689970&$$      28.38&      26.04&      25.53&      25.40& 1&   clump\\						       
       53.145409&      -27.690203&$$      27.94&      25.89&      25.48&      25.40& 2&   clump\\						       
       53.145325&      -27.690252&$$      29.20&      26.06&      25.47&      25.37& 2&   clump\\						       
       53.157550&      -27.690485&$$      26.89&      24.57&      24.28&      24.17& 5&   offcenter clump on disk?\\			       
       53.168747&      -27.690490&$>$      29.85&      26.19&      25.68&     25.52& 2&   double clump\\					       
       53.153418&      -27.691316&$>$      30.09&      26.79&      26.42&     26.13& 4&   several fuzzy\\					       
       53.142828&      -27.691609&$$      29.83&      27.19&      26.98&      27.02& 4&	several fuzzy\\					       
       53.177363&      -27.691677&$>$      29.75&      26.70&      25.74&     25.36& 1&   clump edge on\\					       
       53.143354&      -27.691769&$>$      30.36&      26.87&      26.44&     26.38& 1&   clump\\ 					       
       53.128688&      -27.692143&$$      28.90&      26.83&      26.96&      26.65& 5&   faint clumpy disk?\\				       
       53.132644&      -27.692107&$>$      29.36&      27.10&      27.15&     27.12& 1&   faint clump\\					       
       53.172935&      -27.692512&$>$      30.17&      26.95&      26.30&     26.08& 2&   faint double\\					       
       53.153738&      -27.692797&$$      29.14&      26.74&      25.98&      25.79& 1&   faint clump edge on (tadpole?)\\		       
       53.147447&      -27.693511&$>$      29.98&      26.80&      25.69&     25.38& 2&   double clump\\					       
       53.144209&      -27.693704&$$      29.86&      26.57&      26.24&      26.17& 1&   faint clump\\					       
       53.161386&      -27.694332&$$      27.15&      24.64&      24.31&      24.09& 2&   chain/three clumps\\				       
       53.134128&      -27.694236&$$      29.45&      27.29&      26.53&      26.14& 4&   fuzzy\\						       
       53.168314&      -27.694244&$$      28.74&      26.89&      26.55&      26.03& 4&   fuzzy\\						       
       53.174811&      -27.694860&$$      29.50&      25.51&      25.08&      24.84& 1&   clump (faint tail?)\\				       
       53.159129&      -27.695401&$>$      30.07&      26.50&      26.09&     25.82& 4&   fuzzy\\						       
       53.141777&      -27.695409&$>$      30.19&      26.76&      26.46&     26.39& 4&   faint fuzzy edge on\\				       
       53.176046&      -27.695456&$>$      30.24&      27.22&      26.78&     26.56& 1&   faint clump\\					       
       53.161879&      -27.695566&$>$      29.71&      25.89&      24.99&     24.76& 3&   tadpole\\					       
       53.156940&      -27.695521&$$      30.05&      27.19&      26.77&      26.76& 1&   faint clump close to face on spiral\\		       
       53.134625&      -27.695697&$>$      29.11&      26.02&      25.65&     25.68& 4&   fuzzy\\						       
       53.130320&      -27.695894&$>$      28.80&      25.50&      25.04&     24.81& 3&   tadpole\\					       
       53.166073&      -27.695896&$$      29.41&      25.96&      25.77&      25.76& 1&   fuzzy\\						       
       53.159225&      -27.695951&$$      29.04&      26.95&      26.91&      27.21& 1&   faint clump\\					       
       53.132461&      -27.696248&$>$      29.16&      26.70&      25.76&     25.55& 3&   double clump\\					       
       53.143279&      -27.696576&$$      29.80&      26.99&      26.54&      26.51& 1&   fuzzy close to face on spiral\\			       
       53.138781&      -27.696842&$$      29.28&      26.90&      26.16&      25.60& 1&   fuzzy extended\\				       
       53.133936&      -27.696977&$$      28.51&      26.03&      25.96&      25.55& 1&   fuzzy extended\\				       
       53.151536&      -27.697612&$$      27.68&      25.60&      25.16&      24.87& 3&   tadpole\\					       
       53.136554&      -27.697853&$$      28.24&      27.08&      27.28&      27.36& 4&   fuzzy\\						       
       53.145437&      -27.697987&$$      26.50&      24.67&      24.46&      24.43& 3&tadpole?			\\			       
       53.173859&      -27.698226&$>$      29.78&      26.40&      26.17&     25.65& 4&   faint fuzzy\\					       
       53.152117&      -27.698154&$$      28.77&      26.31&      25.97&      25.86& 3&tadpole?\\						       
       53.174661&      -27.698140&$$      28.40&      26.58&      26.41&      26.39& 4&   fuzzy	\\					       
       53.136690&      -27.698376&$$      28.87&      27.09&      27.04&      27.04& 4&fuzzy			    \\			       
       53.160346&      -27.698494&$$      29.34&      26.98&      26.18&      26.05& 4&   fuzzy\\						       
       53.133898&      -27.698699&$$      27.34&      25.87&      25.71&      25.40& 4&   fuzzy\\						       
       53.149444&      -27.698816&$$      29.62&      27.29&      27.20&      26.96& 1&   faint clump\\					       
       53.182675&      -27.698849&$$      29.16&      26.41&      25.86&      25.59& 1&   clump\\						       
       53.162550&      -27.698872&$$      29.10&      26.68&      26.50&      26.72& 4&   fuzzy, fuzzy companions\\			       
       53.162513&      -27.699074&$$      29.20&      27.03&      26.64&      26.49& 4&   fuzzy, fuzzy companions\\			       
       53.178276&      -27.699784&$$      29.03&      26.95&      26.85&      26.81& 4&   fuzzy with fuzzy companion\\			       
       53.172033&      -27.684078&$$      28.57&      26.31&      25.78&      25.47& 4&   fuzzy\\						       
       53.180185&      -27.699950&$$      27.80&      25.90&      25.69&      25.56& 2&   faint double clump/tadpole?\\			       
       53.137583&      -27.700109&$$      29.01&      26.01&      25.46&      25.09& 1&   clump (spheroid) or faint star?\\		       
       53.169918&      -27.683871&$$      28.82&      26.61&      25.89&      25.60& 4&   fuzzy\\						       
       53.138746&      -27.700467&$>$      29.76&      26.05&      25.26&     25.00& 2&   four clumps\\					       
       53.172890&      -27.683628&$>$      30.20&      27.19&      26.77&     26.32& 1&   faint clump\\					       
       53.173737&      -27.700799&$$      29.27&      26.83&      26.67&      26.91& 4&   fuzzy\\						       
       53.172186&      -27.683993&$>$      30.07&      27.25&      26.77&     26.30& 4&   faint fuzzy, close to clump\\			       
       53.143580&      -27.683502&$>$      28.12&      25.72&      25.40&     25.24& 4&   fuzzy\\						       
       53.169913&      -27.669889&$>$      30.22&      26.60&      26.17&     25.77& 1&   clump\\						       
       53.161684&      -27.682908&$$      28.35&      26.69&      26.70&      26.96& 1&   clump\\						       
       53.163174&      -27.669346&$$      27.54&      25.20&      24.80&      24.67& 5&   clumpy disk\\					       
       53.161053&      -27.672755&$$      29.26&      27.02&      26.41&      26.23& 1&   clump\\						       
       53.140722&      -27.682237&$>$      28.08&      26.79&      26.86&     26.53& 4&   faint fuzzy\\					       
       53.166505&      -27.672539&$$      29.80&      27.19&      26.73&      26.54& 4&   fuzzy\\						       
       53.160749&      -27.671728&$$      27.15&      25.48&      25.23&      25.08& 4&   fuzzy\\						       
       53.164593&      -27.674991&$>$      29.94&      26.48&      25.94&     25.89& 4&   fuzzy\\						       
       53.158301&      -27.680860&$>$      29.15&      26.50&      25.70&     25.33& 4&   fuzzy\\						       
       53.169616&      -27.672793&$$      28.89&      26.71&      26.06&      25.62& 3&   tadpole\\					       
       53.158060&      -27.673707&$$      28.22&      25.78&      25.42&      25.19& 2&   three clumps, chain\\				       
       53.161286&      -27.679136&$$      28.91&      26.94&      27.00&      27.11& 1&   clump\\						       
       53.158157&      -27.673807&$$      27.09&      25.05&      24.83&      24.69& 2&   double clump\\					       
       53.162497&      -27.671140&$$      28.45&      26.32&      25.95&      25.85& 5&   fuzzy edge on disk\\				       
       53.165407&      -27.674053&$$      30.12&      26.92&      26.76&      26.57& 4&   faint fuzzy\\					       
       53.169762&      -27.674803&$$      28.17&      27.06&      27.28&      27.27& 4&   fuzzy\\						       
       53.152429&      -27.677562&$$      27.89&      26.39&      26.25&      26.16& 4&   fuzzy\\						       
       53.156011&      -27.676234&$$      28.80&      25.11&      24.83&      24.80& 3&   tadpole? or peculiar disk\\			       
       53.157810&      -27.676274&$$      28.84&      26.82&      26.62&      26.43& 4&   fuzzy\\                                                      
\enddata
\label{tab:dropoutlist}
\end{deluxetable}


\begin{deluxetable}{l l}
\tablewidth{0pt}
\tablecaption{Morphological classification of $U$ band dropouts made using visual examination of GOODS/ACS $z$-band images. The relative fractions of each class are listed.}
\tablehead{\colhead{Morphological class}& \colhead{Fraction}}
\startdata
Single Clump & 31\%\\
Double or multiple clump& 13\%\\
Clump with tail& 7\%\\
Extended and fuzzy& 38\%\\
Fuzzy, extended with disk-like structure&11\%\\
\enddata
\label{tab:histmorph}
\end{deluxetable}

\clearpage
\end{document}